# GRAVITATIONAL INTRACTION ON QUANTUM LEVEL AND CONSEQUENCES THEREOF

## S.I. Fisenko, I.S. Fisenko

"Rusthermosinthes" JSC,
6 Gasheka Str., 12th Storey,
Moscow 125047
Phone: (+7) 956-82-46, Fax: (+7) 956-82-47
E-mail: StanislavFisenko@yandex.ru

## ABSTRACT

Parity no conservation in the β decay processes is considered as fundamental property of weak interactions. Nevertheless, this property can be treated as anomaly, because in fundamental interactions of the rest types parity is conserved. Analogously, anomaly in the short-duration strong-current pulse discharges is well known. The essence of this phenomenon consists in generation of local high-temperature plasma formations with the typical values of its thermo dynamical parameters exceeding those related to the central section of a discharge. In this paper, an attempt is undertaken to treat these anomalies as manifestation of fundamental properties of gravitational emission. Some consequences of this assumption can be tested in the $\beta$ decay experiments as well as in the experiments with short-duration *z*-pinch-type pulse discharges.

The notion of gravitational emission as an emission of the same level with electromagnetic emission is based on the proven fact of existence of electron's stationary states in its own gravitational field, characterized by gravitational constant $K=10^{42}G$ (G – Newton's gravitational constant).

**1. Gravitational emission of electrons with a banded spectrum as emission of the same level with electromagnetic emission.**

For a mathematical model of interest, which describes a banded spectrum of stationary states of electrons in the proper gravitational field, two aspects are of importance. First. In Einstein's field equations κ is a constant which relates the space-time geometrical properties with the distribution of physical matter, so that the origin of the equations is not connected with the numerical limitation of the κ value. Only the requirement of conformity with the Newtonian Classical Theory of Gravity leads to the small value $\kappa = 8\pi G/c^4$, where $G$, $c$ are, respectively, the Newtonian gravitational constant and the velocity of light. Such requirement follows from the primary concept of the Einstein General Theory of Relativity **(GR)** as a relativistic generalization of the Newtonian Theory of Gravity. Second. The most general form of relativistic gravitation equations are equations with the Λ term. The limiting transition to weak fields leads to the equation

$$\Delta\Phi = -4\pi\rho G + \Lambda c^2,$$

where Φ is the field scalar potential, ρ is the source density. This circumstance, eventually, is crucial for neglecting the Λ term, because only in this case the GR is a generalization of the Classical Theory of Gravity. Therefore, the numerical values of $\kappa = 8\pi G/c^4$ and $\Lambda = 0$ in the GR equations are not associated with the origin of the equations, but follow only from the conformity of the GR with the classical theory.

From the 70's onwards, it became obvious [1] that in the quantum region the numerical value of $G$ is not compatible with the principles of quantum mechanics. In a number of papers [1] (including also [2]) it was shown that in the quantum region the coupling constant $K$ ($K \approx 10^{40} G$) is acceptable. The essence of the problem of the generalization of relativistic equations on the quantum level was thus outlined: such generalization must match the numerical values of the gravity constants in the quantum and classical regions.

In the development of these results, as a micro-level approximation of Einstein's field equations, a model is proposed, based on the following assumption:

"*The gravitational field within the region of localization of an elementary particle having a mass $m_0$ is characterized by the values of the gravity constant K and of the constant Λ that lead to the stationary states of the particle in its proper gravitational field, and the particle stationary states as such are the sources of the gravitational field with the Newtonian gravity constant G*".

The most general approach in the Gravity Theory is the one which takes twisting into account and treats the gravitational field as a gage field, acting on equal terms with other fundamental fields ([3]). Such approach lacks in apriority gives no restrictions on the microscopic level. For an elementary spinor source with a mass $m_0$, the set of equations describing its states in the proper gravitational field in accordance with the adopted assumption will have the form

$$\{i\gamma^{\mu}(\nabla_{\mu} + \bar{\kappa}\overline{\Psi}\gamma_{\mu}\gamma_5\Psi\gamma_5) - m_0 c/\hbar\}\Psi = 0 \tag{1}$$

$$R_{\mu\nu} - \frac{1}{2}g_{\mu\nu}R = -\kappa\{T_{\mu\nu}(E_n) - \mu g_{\mu\nu} + (g_{\mu\nu}S_\alpha S^\alpha - S_\mu S_\nu)\} \tag{2}$$

$$R(K, \Lambda, E_n, r_n) = R(G, E'_n, r_n) \tag{3}$$

$$\{i\gamma^{\mu}\nabla_{\mu} - m_n c/\hbar\}\Psi' = 0 \tag{4}$$

$$R_{\mu\nu} - \frac{1}{2}g_{\mu\nu}R = -\kappa' T_{\mu\nu}(E'_n) \tag{5}$$

The following notations are used throughout the text of the paper: $\kappa = 8\pi K/c^4$, $\kappa' = 8\pi G/c^4$, $E_n$ is the energy of stationary states in the proper gravitational field with the constant $K$, $\Lambda = \kappa\mu$, $r_n$ is the value of the coordinate $r$ satisfying the equilibrium n-state in the proper gravitational field, $\bar{\kappa} = \kappa_0\kappa$, $\kappa_0$ is the dimensionality constant, $S_a = \overline{\Psi}\gamma_a\gamma_5\Psi$, $\nabla_\mu$ is the spinor-coupling covariant derivative independent of twisting, $E'_n$ is the energy state of the particle having a mass $m_n$ (either free of field or in the external field), described by the wave function $\psi'$ in the proper gravitational field with the constant $G$. The rest of the notations are generally adopted in the gravitation theory.

Equations (1) through (5) describe the equilibrium states of particles (stationary states) in the proper gravitational field and define the localization region of the field characterized by the constant $K$ that satisfies the equilibrium state. These stationary states are sources of the field with the constant $G$, and condition (3) provides matching the solution with the gravitational constants $K$ and $G$. The proposed model in the physical aspect is compatible with the principles of quantum mechanics principles, and the gravitational field with the constants $K$ and $\Lambda$ at a certain, quite definite distance specified by the equilibrium state transforms into the filed having the constant $G$ and satisfying, in the weak field limit, the Poisson equation.

The set of equations (1) through (5), first of all, is of interest for the problem of stationary states, i.e., the problem of energy spectrum calculations for an elementary source in the own gravitational field. In this sense it is reasonable to use an analogy with electrodynamics, in particular, with the problem of electron stationary states in the Coulomb field. Transition from the Schrödinger equation to the Klein-Gordon relativistic equations allows taking into account the fine structure of the electron energy spectrum in the Coulomb field, whereas transition to the Dirac equation allows taking into account the relativistic fine structure and the energy level splitting associated with spin-orbital interaction. Using this analogy and the form of equation (1), one can conclude that solution of this equation without the term $\kappa \overline{\Psi} \gamma_\mu \gamma_5 \Psi \gamma_5$ gives a spectrum similar to that of the fine structure (similar in the sense of relativism and removal of the principal quantum number deµgeneracy).. Taking the term $\kappa \overline{\Psi} \gamma_\mu \gamma_5 \Psi \gamma_5$ into account, as is noted in [1], is similar to taking into account of the term $\overline{\Psi} o^{\mu\nu} \Psi F_{\mu\nu}$ in the Pauli equation. The latter implies that the solution of the problem of stationary states with twisting taken into account will give a total energy-state spectrum with both the relativistic fine structure and energy state splitting caused by spin-twist interaction taken into account. This fact, being in complete accord with the requirements of the Gauge Theory of Gravity, allows us to believe that the above-stated assumptions concern ing the properties of the gravitational field in the quantum region are relevant, in the general case, just to the gravitational field with twists.

Complexity of solving this problem compels us to employ a simpler approximation, namely,: energy spectrum calculations in a relativistic fine-structure approximation. In this approximation the problem of the stationary states of an elementary source in the proper gravitational field well be reduced to solving the following equations:

$$f'' + \left(\frac{v' - \lambda'}{2} + \frac{2}{r}\right)f' + e^{\lambda}\left(K_n^2 e^{-v} - K_0^2 - \frac{l(l+1)}{r^2}\right)f = 0 \tag{6}$$

$$-e^{-\lambda}\left(\frac{1}{r^2} - \frac{\lambda'}{r}\right) + \frac{1}{r^2} + \Lambda = \beta(2l+1)\left\{f^2\left[e^{-\lambda}K_n^2 + K_0^2 + \frac{l(l+1)}{r^2}\right] + f'^2 e^{-\lambda}\right\} \tag{7}$$

$$-e^{-\lambda}\left(\frac{1}{r^2} + \frac{v'}{r}\right) + \frac{1}{r^2} + \Lambda = \beta(2l+1)\left\{f^2\left[K_0^2 - K_n^2 e^{-v} + \frac{l(l+1)}{r^2}\right] - e^{\lambda}f'^2\right\} \tag{8}$$

$$\left\{-\frac{1}{2}(v'' + v'^2) - (v' + \lambda')\left(\frac{v'}{4} + \frac{1}{r}\right) + \frac{1}{r^2}(1 + e^{\lambda})\right\}_{r=r_n} = 0 \tag{9}$$

$$f(0) = const \ll \infty \tag{10}$$

$$f(r_n) = 0 \tag{11}$$

$$\lambda(0) = v(0) = 0 \tag{12}$$

$$\int_0^{r_n} f^2 r^2 dr = 1 \tag{13}$$

Equations (6)—(8) follow from equations (14)—(15)

$$\left\{-g^{\mu\nu}\frac{\partial}{\partial x_\mu}\frac{\partial}{\partial x_\nu} + g^{\mu\nu}\Gamma^\alpha_{\mu\nu}\frac{\partial}{\partial x_\alpha} - K_0^2\right\}\Psi = 0 \tag{14}$$

$$R_{\mu\nu} - \frac{1}{2}g_{\mu\nu}R = -\kappa(T_{\mu\nu} - \mu g_{\mu\nu}), \tag{15}$$

after the substitution of $\Psi$ in the form: $\Psi = f_{El}(r)Y_{lm}(\theta, \varphi)\exp\left(\frac{-iEt}{\hbar}\right)$ into them and

5   specific computations in the central-symmetry field metric with the interval defined by the expression [4]

$$dS^2 = c^2 e^v dt^2 - r^2(d\theta^2 + \sin^2\theta d\varphi^2) - e^\lambda dr^2 \tag{16}$$

The following notation is used above: $f_m$ is the radial wave function that describes the states with a definite energy $E$ and the orbital moment $l$ (hereafter the subscripts $El$ are

10   omitted), $Y_{lm}(\theta, \varphi)$ - are spherical functions, $K_n = E_n/\hbar c$, $K_0 = cm_0/\hbar$, $\beta = (\kappa/4\pi)(\hbar/m_0)$.

Condition (9) defines $r_n$, whereas equations (10) through (12) are the boundary conditions and the normalization condition for the function $f$, respectively. Condition (9) in the general case has the form $R(K, r_n) = R(G, r_n)$. Neglecting the proper gravitational field with the constant $G$, we shall write down this condition as $R(K, r_n) = 0$, to which

15   equality (9) actually corresponds.

The right-hand sides of equations (7)—(8) are calculated basing on the general expression for the energy-momentum tensor of the complex scalar field:

$$T_{\mu\nu} = \Psi^+_{,\mu}\Psi_{,\nu} + \Psi^+_{,\nu}\Psi_{,\mu} - \left(\Psi^+_{,\mu}\Psi^{,\mu} - K_0^2\Psi^+\Psi\right) \tag{17}$$

The appropriate components $T_{\mu\nu}$ are obtained by summation over the index $m$ with application of characteristic identities for spherical functions [5] after the substitution of $\Psi = f(r)Y_{lm}(\theta,\varphi)\exp\left(\dfrac{-iEt}{\hbar}\right)$ into (17).

Even in the simplest approximation the problem of the stationary states of an elementary source in the proper gravitational field is a complicated mathematical problem. It becomes simpler if we confine ourselves to estimating only the energy spectrum. Equation (6) can be reduced in many ways to the equations [6]

$$f' = fP(r) + Q(r)z \qquad\qquad z' = fF(r) + S(r)z \tag{18}$$

This transition implies specific choice of $P, Q, F, S$, such that the conditions

$$P + S + Q'/Q + g = 0 \qquad\qquad FQ + P' + P^2 + Pg + h = 0 \tag{19}$$

should be fulfilled, where $g$ and $h$ correspond to equation. (6) written in the form: $f'' + gf'$ + $hf = Q$. Conditions (19) are satisfied, in particular, by $P, Q,, F, S$ written as follows:

$$Q = 1, \quad P = S = -g/2, \quad F = \frac{1}{2}g' + \frac{1}{4}g^2 - h \tag{20}$$

Solutions of set (18) will be the functions: [6]

$$f = C\rho(r)\sin\theta(r) \qquad\qquad z = C\rho(r)\cos\theta(r) \tag{21}$$

where $C$ is an arbitrary constant, $\theta(r)$ is the solution of the equation:

$$\theta' = Q\cos^2\theta + (P - S)\sin\theta\cos\theta - F\sin^2\theta, \tag{22}$$

and $\rho(r)$ is found from the formula

$$\rho(r) = \exp\int_0^r\left[P\sin^2\theta + (Q + F)\sin\theta\cos\theta + S\cos^2\theta\right]dr. \tag{23}$$

In this case, the form of presentation of the solution in polar coordinates makes it possible to determine zeros of the functions $f(r)$ at $r = r_n$, with corresponding values of $\theta = n\pi$ ($n$ being an integer). As one of the simplest approximations for $v, \lambda$, we shall choose the dependence:

$$e^v = e^{-\lambda} = 1 - \frac{\tilde{r}_n}{r + C_1} + \Lambda(r - C_2)^2 + C_3 r \tag{24}$$

where

$$\tilde{r}_n = \frac{2Km_n}{c^2} = \frac{2KE_n}{c^4} = \left(\frac{2K\hbar}{c^3}\right)K_n, \quad C_1 = \frac{\tilde{r}_n}{\Lambda r_n^2}, \quad C_2 = r_n, \quad C_3 = \frac{\tilde{r}_n}{r_n(r_n + C_1)}$$

Earlier the estimate for $K$ was adopted to be $K \approx 1.7 \times 10^{29}$ Nm$^2$kg$^{-2}$. If we assume that the observed value of the electron rest mass $m_1$ is its mass in the ground stationary state in the proper gravitational field, then $m_0 = 4m_1/3$. From dimensionality considerations it follows that energy in the bound state is defined by the expression $\left(\sqrt{Km_0}\right)^2 / r_1 = 0.17 \times 10^6 \times 1.6 \times 10^{-19}$ J, where $r_1$ is the classical electron radius. This leads to the estimate $K \approx 5.1 \times 10^{31}$ Nm$^2$kg$^{-2}$ which is later adopted as the starting one. It is known that by use of the dependence $E_0 = \dfrac{e^2}{r_0} = mc^2$ [7] for the impulse we receive the expression $P_i = \dfrac{4}{3} \dfrac{E_0}{c^2} v_i$, that is differentiated by the multiplier 4/3 from the correct expression for the impulse of the particles, the mass of which is $m = \dfrac{E_0}{c^2}$. It is this fact that points at the correctness of the estimates received for the electron, because the lacking part of the energy is in the bound state. Discrepancies in the estimates $K$ obtained by different methods are quite admissible, all the more so since their character is not catastrophic. From the condition that $\mu$ is the electron energy density it follows: $\mu = 1.1 \times 10^{30}$ J/m$^3$, $\Lambda = \kappa\mu = 4.4 \times 10^{29}$ m$^{-2}$. From (22) (with the equation for $f(r)$ taken into account) it follows:

$$2\theta' = (1-\overline{F}) + (1+\overline{F})\cos 2\theta \approx (1-\overline{F}), \qquad (25)$$

where

$$\overline{F} = \frac{1}{2}\overline{g}' + \frac{1}{4}\overline{g}^2 - \overline{h}, \quad \overline{g} = r_n\left(\frac{2}{r} + \frac{(v'-\lambda')}{2}\right), \quad \overline{h} = r_n^2 e^\lambda\left(K_n^2 e^{-v} - K_0^2 - \frac{l(l+1)}{r^2}\right).$$

The integration of equation (25) and substitution of $\theta = \pi n$, $r = r_n$ give the relation between $K_n$ and $r_n$:

$$-2\pi n = -\frac{7}{4} - \frac{r_n K_n^2}{\Lambda^2}\sum_{i=1}^{3}\left\{A_i\left[\frac{(r_n+\alpha_i)^2}{2} - 2\alpha_i(r_n+\alpha_i) + \frac{\alpha_i^3}{(r_n+\alpha_i)} + 2C_1(r_n+\alpha_i) + \right.\right.$$
$$\left.+2C_1\frac{\alpha_i^2}{r_n+\alpha_i} + \frac{C_2^2\alpha_i}{r_n+\alpha_i}\right] + B_i\left[(r_n+\alpha_i) + \alpha_i^2\frac{1}{r_n+\alpha_i} + \frac{2C_1\alpha_i}{r_n+\alpha_i} - \frac{C_2^2}{r_n+\alpha_i}\right]\right\} +$$
$$+\frac{K_0^2 r_n}{\Lambda^2}\sum_{i=1}^{3}A_i'(r_n+\alpha_i) + \frac{r_n l(l+1)}{\Lambda}\left[d_1 r_n - \frac{C_1 d_2}{r_n} + \sum_{i=1}^{3}a_i(r_n+\alpha_i)\right] - \qquad (26)$$
$$-\frac{K_n^2 r_n}{\Lambda^2}\left\{\sum_{i=1}^{3}\left[2\alpha_i^2 A_i - 2\alpha_i B_i - 4C_1 A_i \alpha_i + 2C_1 B_i + C_2^2 A_i + \frac{K_0^2 \Lambda A_i'}{K_n^2}(\alpha_i - C_1) - \right.\right.$$
$$\left.\left.- r_n^2 \Lambda l(l+1)a_i(C_1-\alpha_i)\right]\ln(r_n+\alpha_i) - r_n \Lambda^{-1} l(l+1)(d_2+C_1 d_1)\ln r_n\right\}$$

The coefficients entering into equation (26) are coefficients at simple fractions in the expansion of polynomials, required for the integration, wherein $\alpha_i \sim K_n$, $d_2 \sim A_i \sim r_n^{-5}$, $B_i \sim r_n^{-4}$, $A'_i \sim r_n^{-2}$, $a_i \sim r_n^{-4}$, $d_1 = r_n^{-4}$. For eliminating $r_n$ from (26), there exists condition (9) (or the condition $\exp \nu(K,r_n) = 1$ equivalent to it for the approximation employed), but its direct use will complicate the already cumbersome expression (26) still further. At the same time, it easy to note that $r_n \sim 10^{-3} r_{nc}$, where $r_{nc}$ is the Compton wavelength of a particle of the mass $m_n$, and, hence, $r_n \sim 10^{-3} K_n^{-1}$. The relation (26) per se is rather approximate, but, nevertheless, its availability, irrespective of the accuracy of the approximation, implies the existence of an energy spectrum as a consequence of the particle self-interaction with its own gravitational field in the range $r \leq r_n$, where mutually compensating action of the field and the particle takes place. With $l = 0$ the approximate solution (26), with the relation between $r_n$ and $K_n$ taken into account, has the form

$$E_n = E_0 \left(1 + \alpha e^{-\beta n}\right)^{-1}, \tag{27}$$

where $\alpha = 1.65$, $\beta = 1.60$.

The relation (27) is concretized, proceeding from the assumption that the observed value of the electron rest mass is the value of its mass in the grounds stationary state in the proper gravitational field, the values $r_1 = 2.82 \times 10^{-15}$ m, $K_l = 0.41 \times 10^{12}$ m$^{-1}$ giving exact zero of the function by the very definition of the numerical values for $K$ and $\Lambda$.

So, the presented numerical estimates for the electron show that within the range of its localization, with $K \sim 10^{31}$ N m$^2$ kg$^{-2}$ and $\Lambda \sim 10^{29}$ m$^{-2}$, there exists the spectrum of stationary states in the proper gravitational field. The numerical value of $K$ is, certainly, universal for any elementary source. *Existence of such numerical value $\Lambda$ denotes a phenomenon having a deep physical sense: introduction into density of the Lagrange function of a constant member independent on a state of the field. This means that the time-space has an inherent curving which is connected with neither the matter nor the gravitational waves.* The distance at which the gravitational field with the constant K is localized is less than the Compton wavelength, and for the electron, for example, this value is of the order of its classical radius. At distances larger than this one, the gravitational field is characterized by the constant $G$, i.e., correct transition to Classical GTR holds.

From equation (27) there follow in a rough approximation the numerical values of the stationary state energies: $E_1 = 0.511$ MeV, $E_2 = 0.638$ MeV, ... $E_\infty = 0.681$ MeV.

The existence of stationary states in own gravitational field also completely corresponds the special relativity theory. According to SRT, relativistic link between energy and impulse is broken, if we assume that full electron's energy is defined only by Lawrence's electromagnetic energy [7]. If to expand the situation it is as follows [7]. Energy and impulse of the moving electron (with the assumption that the distribution of the electric charge is spherically symmetrical) is defined by expression:

$$\overline{P} = \overline{V} \frac{\frac{4}{3} E_0 / c^2}{\sqrt{1-\beta^2}} \qquad (28)$$

$$E = \frac{E_0(1 + \frac{1}{3} V^2 / c^2)}{\sqrt{1-\beta^2}} \qquad (29)$$

If these expressions were at the same time defining full impulse and full energy, the following relator would take place:

$$E = \int (\overline{V} \frac{d\overline{P}}{dt}) dt \qquad (30)$$

However this relator cannot take place as integral in the right part equals to:

$$\frac{\frac{4}{3} E_0}{\sqrt{1-\beta^2}} + const \qquad (31)$$

If we find that impulse contrary to the energy has strictly electromagnetic character, then to $E'$ of the moving and full energy $E_0'$ of the resting electron, and also to rest mass $E_0$, following relators will take place:

$$E' = \frac{E_0'}{\sqrt{1-\beta^2}}, \ E_0' = \frac{4}{3} E_0, \ m_0 = \frac{E_0'}{c^2} = \frac{4}{3} \frac{E_0}{c^2}, \qquad (32)$$

Where rest mass $m_0$ is defined by following expression:

$$\overline{P} = \frac{m_0 \overline{V}}{\sqrt{1-\beta^2}} \qquad (33)$$

Then from (32) it follows that full energy of resting electron equals to $\frac{4}{3}$ of its Lawrence's electromagnetic energy. Numeric data of the electron's stationary states spectrum in own gravitational field fully correspond to it.

Quantum transitions over stationary states must lead to the gravitational emission characterized by the constant $K$ with transition energies starting from 127 keV to 170 keV Two circumstances are essential here.

*First.* The correspondence between the electromagnetic and gravitational interaction takes place on replacement of the electric charge e by the gravitational "charge" m$\sqrt{K}$, so that the numerical values K place the electromagnetic and gravitational emission effects on the same level (for instance, the electromagnetic and gravitational bremsstrahlung cross-sections will differ only by the factor 0.16 in the region of coincidence of the emission spectra).

*Second.* The natural width of the energy levels in the above-indicated spectrum of the electron stationary states will be very small. The small value of the energy level widths, compared to the electron energy spread in real conditions, explains why the gravitational emission effects are not observed as a mass phenomenon in epiphenomena, e.g., in the processes of electron beam bremsstrahlung on targets. However, there is a possibility of registration of gravitational emission spectrum lines. The results are given below. There is certain analytic interest in β-decay processes with asymmetry of emitted electrons [8], due to (as it is supposed to be) parity violation in weak interactions. β - asymmetry in angular distribution of electrons was registered for the first time during experiments with polarized nucleuses $_{27}Co^{60}$, β-spectrum of which is characterized by energies of MeV. If in the process of β-decay exited electrons are born, then along with decay scheme

$$n \rightarrow p + e^- + \tilde{v} \tag{34}$$

there will be also decay scheme

$$n \rightarrow p + (e^*)^- + \tilde{v} \rightarrow e^- + \tilde{\gamma} + \tilde{v} \tag{35}$$

where $\tilde{\gamma}$ is graviton.

Decay (35) is energetically limited by energy values of 1 MeV order (in rough approximation), taking into consideration that the difference between lower excitation level of electron's energy (in own gravitational field) and general <100 keV and the very character β-spectrum. Consequently, $_{27}Co^{60}$ nucleuses decay can proceed with equal probability as it is described in scheme (34) or in scheme (35). For the light nucleuses, such as $_1H^3$ β-decay can only proceed as it is described in scheme (34). At the same time, emission of graviton by electron in magnetic field can be exactly the reason for β-asymmetry in angular distribution of electrons. If so, then the phenomenon of β-asymmetry will not be observed in light β-radioactive nucleuses. This would mean that β-asymmetry in angular distribution of electrons, which is interpreted as parity violation, is the result of electron's gravitational emission, which should be manifested in existence of lower border β-decay, as that's where β-asymmetry appears to be.

## 2. Gravitational Emission in Dense High-Temperature Plasma
### 2.1. Excitation of Gravitational Emission in Plasma

For the above-indicated energies of transitions over stationary states in the own field and the energy level widths, the sole object in which gravitational emission can be realized as a mass phenomenon will be, as follows from the estimates given below, a dense high-temperature plasma.

Using the Born approximation for the bremsstrahlung cross-section, we can write down the expression for the electromagnetic bremsstrahlung per unit of volume per unit of time as

$$Q_e = \frac{32}{3} \frac{z^2 r_0^2}{137} mc^2 \, n_e \, n_i \, \frac{\sqrt{2k T_e}}{\pi m} = 0.17 \times 10^{-39} z^2 n_e n_i \sqrt{T_e}, \quad (36)$$

where $T_e$, $k$, $n_i$, $n_e$, $m$, $z$, $r_o$ are the electron temperature, Boltzmann's constant, the concentration of the ionic and electronic components, the electron mass, the serial number of the ionic component, the classical electron radius, respectively.

Replacing $r_o$ by $r_g = 2K\,m/c^2$ (which corresponds to replacing the electric charge $e$ by the gravitational charge $m\sqrt{K}$), we can use for the gravitational bremsstrahlung the relation

$$Q_g = 0.16 Q_e. \quad (37)$$

From (36) it follows that in a dense high-temperature plasma with parameters $n_e = n_i = 10^{23}$ m$^{-3}$, $T_e = 10^7$ K, the specific power of the electromagnetic bremsstrahlung is equal to $\approx 0.53\,10^{10}$ J/m$^3$ s, and the specific power of the gravitational bremsstrahlung is $0.86\,10^9$ J/m$^3$ s. These values of the plasma parameters, apparently, can be adopted as guide threshold values of an appreciable gravitational emission level, because the relative proportion of the electrons whose energy on the order of the energy of transitions in the own gravitational field, diminishes in accordance with the Maxwellian distribution exponent as $T_e$ decreases.

### 2.2. Amplification of Gravitational Emission in Plasma

For the numerical values of the plasma parameters $T_e = T_i = (10^7 - 10^8)$K, $n_e = n_i = (10^{23} - 10^{25})$ m$^{-3}$ the electromagnetic bremsstrahlung spectrum will not change essentially with Compton scattering of electron emission, and the bremsstrahlung itself is a source of emission losses of a high-temperature plasma. The frequencies of this continuous spectrum are on the order of $(10^{18}$—$10^{20})$ s$^{-1}$, while the plasma frequency for the above-cited plasma parameters is $(10^{13}$—$10^{14})$ s$^{-1}$, or 0.1 eV of the energy of emitted quanta.

*The fundamental distinction of the gravitational bremsstrahlung from the electromagnetic bremsstrahlung is the banded spectrum of the gravitational emission, corresponding to the spectrum of the electron stationary states in the own gravitational field.*

The presence of cascade transitions from the upper excited levels to the lower ones will lead to that the electrons, becoming excited in the energy region above 100 keV, will be emitted, mainly, in the eV region, i.e., energy transfer along the spectrum to the low-frequency region will take place. Such energy transfer mechanism can take place only in quenching spontaneous emission from the lower electron energy levels in the own gravitational field, which rules out emission with quantum energy in the keV region. A detailed description of the mechanism of energy transfer along the spectrum will hereafter give its precise numerical characteristics. Nevertheless, undoubtedly, the very fact of its existence, conditioned by the banded character of the spectrum of the gravitational bremsstrahlung, can be asserted. The low-frequency character of the gravitational bremsstrahlung spectrum will lead to its amplification in plasma by virtue of the locking condition $\omega_g \leq 0.5\sqrt{10^3 n_e}$ being fulfilled.

From the standpoint of practical realization of the states of a high-temperature plasma compressed by the emitted gravitational field, two circumstances are of importance.

*First.* Plasma must comprise two components, with multiply charged ions added to hydrogen, these ions being necessary for quenching spontaneous emission of electrons from the ground energy levels in the own gravitational field. For this purpose it is necessary to have ions with the energy levels of electrons close to the energy levels of free excited electrons. Quenching of the lower excited states of the electrons will be particularly effective in the presence of a resonance between the energy of excited electron and the energy of electron excitation in the ion (in the limit, most favorable case — ionization energy). An increase of z increases also the specific power of the gravitational bremsstrahlung, so that on the condition $\omega_g \leq 0.5\sqrt{10^3 n_e}$ being fulfilled, the equality of the gas-kinetic pressure and the radiation pressure

$$k(n_e T_e + n_i T_i) = 0.16(0.17 \; 10^{-39} z^2 n_e n_i \sqrt{T_e}) \Delta t \qquad (38)$$

will take place at $\Delta t = (10^{-6} - 10^{-7})$ s for the permissible parameter values of compressed plasma $n_e = (1 + a) n_i = (10^{25} - 10^{26})$ m$^{-3}$, $a > 2$, $T_e \approx T_e = 10^8$ K, $z > 10$.

*Second.* The necessity of plasma ejection from the region of the magnetic field with the tentative parameters $n_e = (10^{23} - 10^{24})$ m$^{-3}$, $T_e = (10^7 - 10^8)$ K with subsequent energy pumping from the magnetic field region.

**2.3. A series of actions required for obtaining steady states of dense-high temperature plasma**

- Forming and accelerating binary plasma with multivalent ions by accelerating magnetic field in a pulse high-current discharge.

- Injection of binary plasma from the space of the accelerating magnetic field:

exciting stationary states of an electron in its own gravitational field in the range of energy up to 171 keV with following radiation (Fig. 1) under the condition of quenching lower excited energy levels of ion electron shell of a heavy component (Fig. 2, including quenching excited state of electrons directly in nuclei of small sequential number as carbon) when retarding plasma bunch ejected from the space of the accelerating magnetic field. Cascade transitions from the upper levels are realized in the process of gravitational radiation energy transit to long-wave range.

The sequence of the operations is carried out in a two-sectional chamber (Fig. 3); the structure of the installation is most suitable for the claimed method of forming steady states of the dense high-temperature plasma [9]) with magnetodynamic outflow of plasma and further conversion of the plasma bunch energy (in the process of quenching) in the plasma heat energy for securing both further plasma heating and exciting gravitational radiation and its transit into a long-wave part of the spectrum with consequent plasma compression in the condition of radiation blocking and increasing.

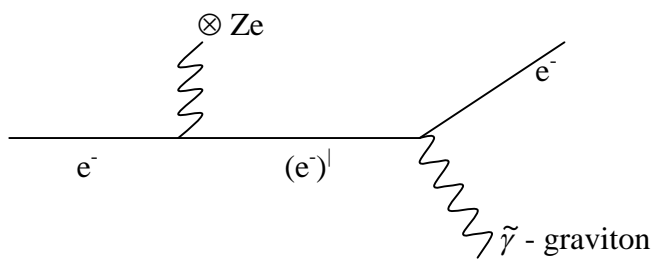

Figure 1. Graviton emission when quenching an electron in a nucleus.

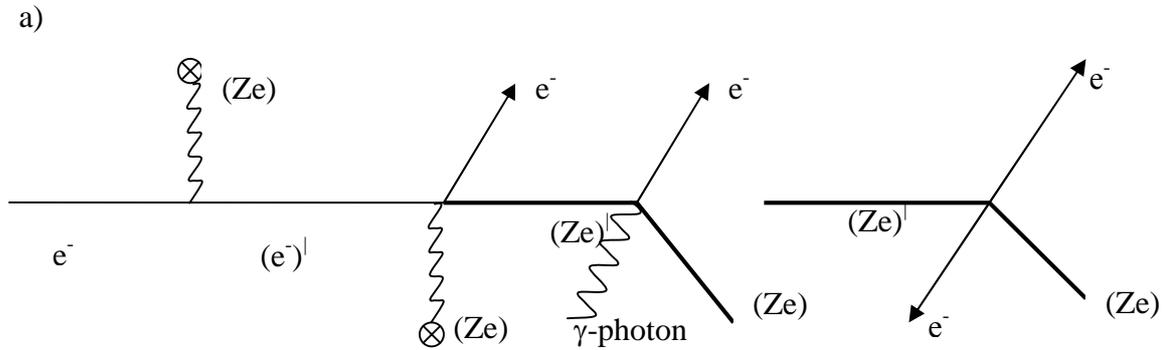

Photoeffect     Auger Effect

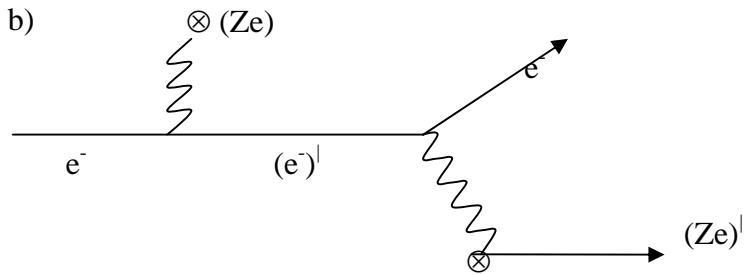

Figure 2. Quenching lower excited states of electron by: a) many-electron ions (photoelectric effect with release of one electron or autoionization (Auger effect) with release of two electrons depending on the ion number and quenching energy); b) nuclei without electron shells when an excited electron returns to normal state transferring excess energy directly to the nucleus with higher probability for the lover energy levels of excited electrons.

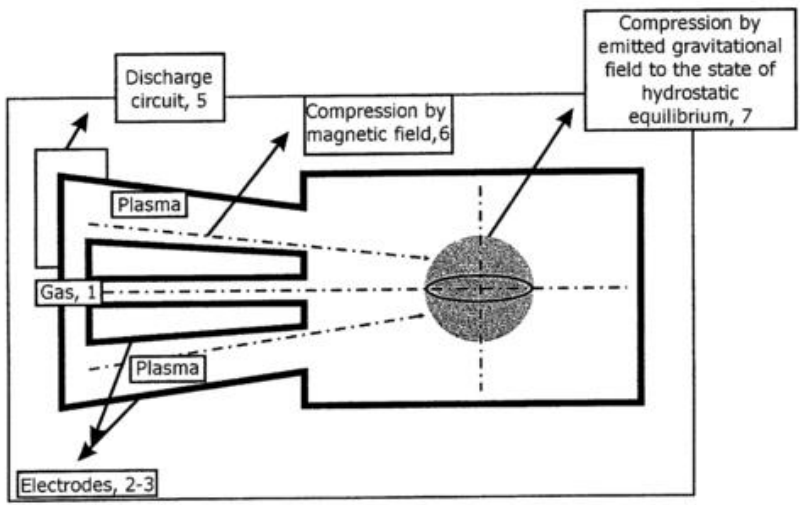

Figure 3. Forming of stable states of dense high-temperature plasma: the scheme

Of interest there are two modes of the installation operations depending on the work gas composition:

- a composition with hydrogen and xenon providing only for achieving steady states of plasma with consequent realization of thermonuclear reactions for compositions of (d+t) + multi-charge atoms type;

- a composition with hydrogen and carbon providing thermonuclear reactions of carbon cycle in plasma steady state mode, including energy pick-up in the form of electromagnetic radiation energy.

## 3. Experimental data

### 3.1. Registration of electron gravitational radiation lines and energy spectrum in their own gravitational field:

It is known that the form of free neutron decay β-spectrum satisfactorily corroborates theoretical dependence for allowed transitions except soft parts of β-spectrum. Corresponding theoretical and experimental spectra are shown in Figs. 4, 5. The soft part of the spectrum is clearly linear exactly corresponding (taking into account kinetic energy of an outgoing electron) to the spectrum of electron steady states in its own gravitational field in then range of the steady state energies up to 171 keV.

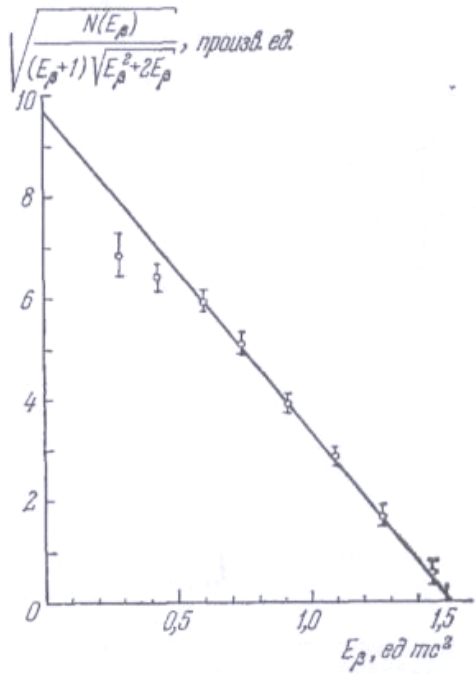 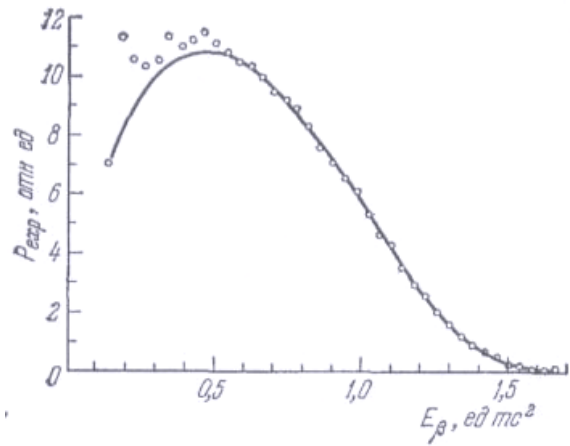

Figure 4. Beta-spectrum of free neutron decay obtained by Robson [10].

The strait line is Fermi graph, the experimental data points according to Robson [10]

Figure 5. Beta-spectrum of free neutron decay obtained by Christensen et al. [11].

The curve corresponds to a theoretical spectrum corrected for spectrometer energy resolution.

In independent experiments [12] when at the same time electron energy distribution after electron beam passing through a foil was registered, clearly line energy spectrum was observed: Fig. 6(a). The line radiation spectrum is also clearly seen: Fig. 6(b) which can not be explained only by the presence of accelerated electron groups. The quantitative identification of the spectrum requires more precise and broad measurements including identification algorithm of energy spectrum quantitative values relating directly to steady states of electrons. Nevertheless, registered the line type of electron energy spectrum and corresponding line electron radiation spectrum preliminary corroborate as a rough approximation the very fact of electron steady states in their own gravitational field exactly in the energy range up to 171 keV.

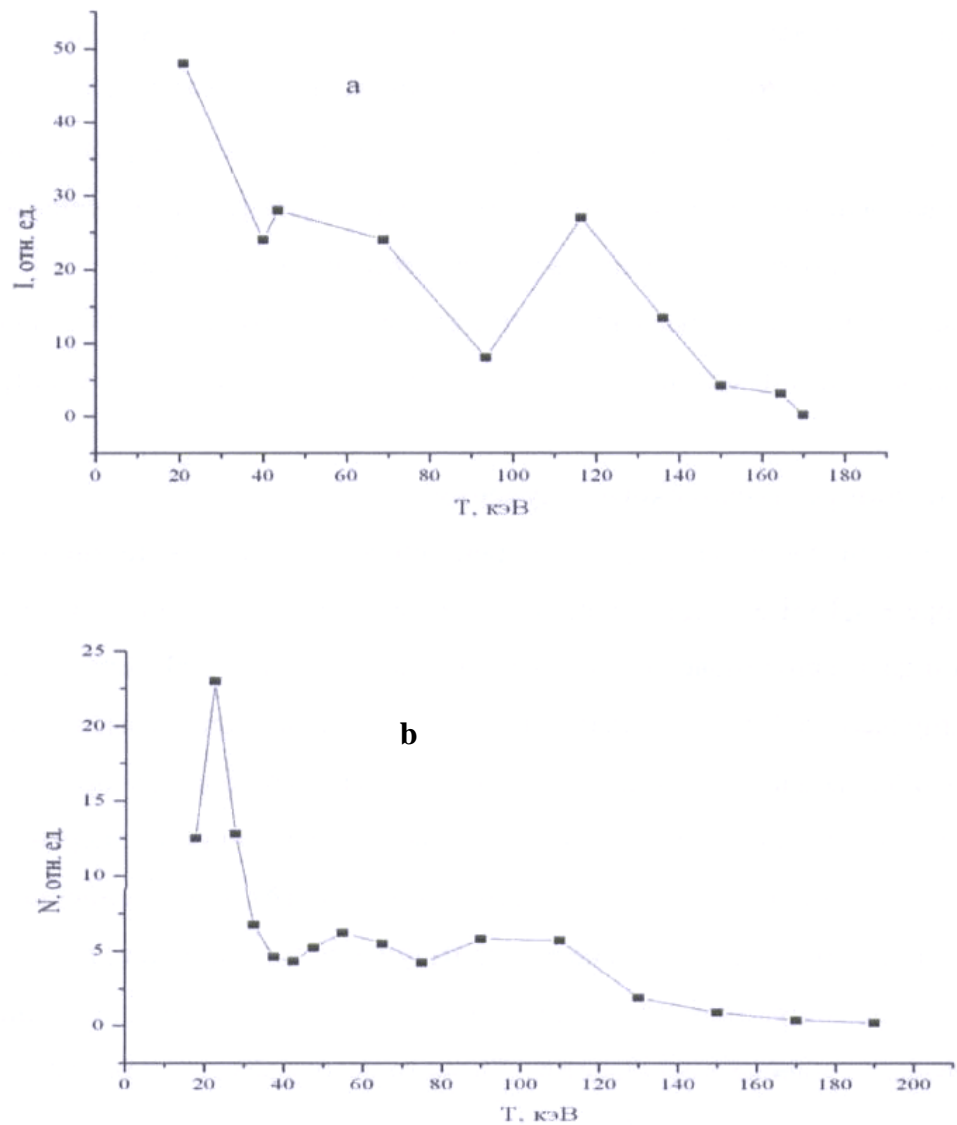

Figure 6. Energy distribution of (a) electrons and (b) X-ray quanta [12]

It is obvious that these data need to be supplemented with direct experimental identification both regarding both electron gravitational radiation spectrum lines and electron steady state energy spectrum in its own gravitational field. Fig. 7 shows electron beam energy spectra in a pulse accelerator measured by a semicircular magnetic spectrometer. Two peaks of the energy spectra are connected to the feature of the pulse accelerator operations, the secondary pulse is due to lower voltage. This leads to the second (low-energy) maximum of the energy spectrum distribution.

A telemetry error in the middle and soft parts of the spectrum is not more than ± 2%. The magnetic spectrometer was used for measuring the energy spectrum of electrons after passing through the accelerator anode grid and also spectra of electrons after passing though a foil arranged behind the accelerator mesh anode. These data (and the calculated spectrum) are presented in Fig. 7. Similar measurements were carried out for Ti foil (foil thickness 50 μm) and Ta (foil thickness 10 μm). In case of Ti the measurements were limited from the top by energy of 0.148 MeV, and in case of TA by energy of 0.168 MeV. Above these values the measurement errors increase substantially (for this type of the accelerator). The difference between the normalized spectral densities of theoretical and experimental electron spectra after passing through Ti, Ta and Al foils is shown on Fig.9. The data indicate that there is a spectrum of electron energy states in their own gravitational field when the electrons are excited when passing through a foil. The obtained data are not sufficient for numerical spectrum identification but the very fact of the spectrum presence according to the data is doubtless.

$\frac{\Delta N_e}{\Delta E}, \frac{electron}{MeV}$

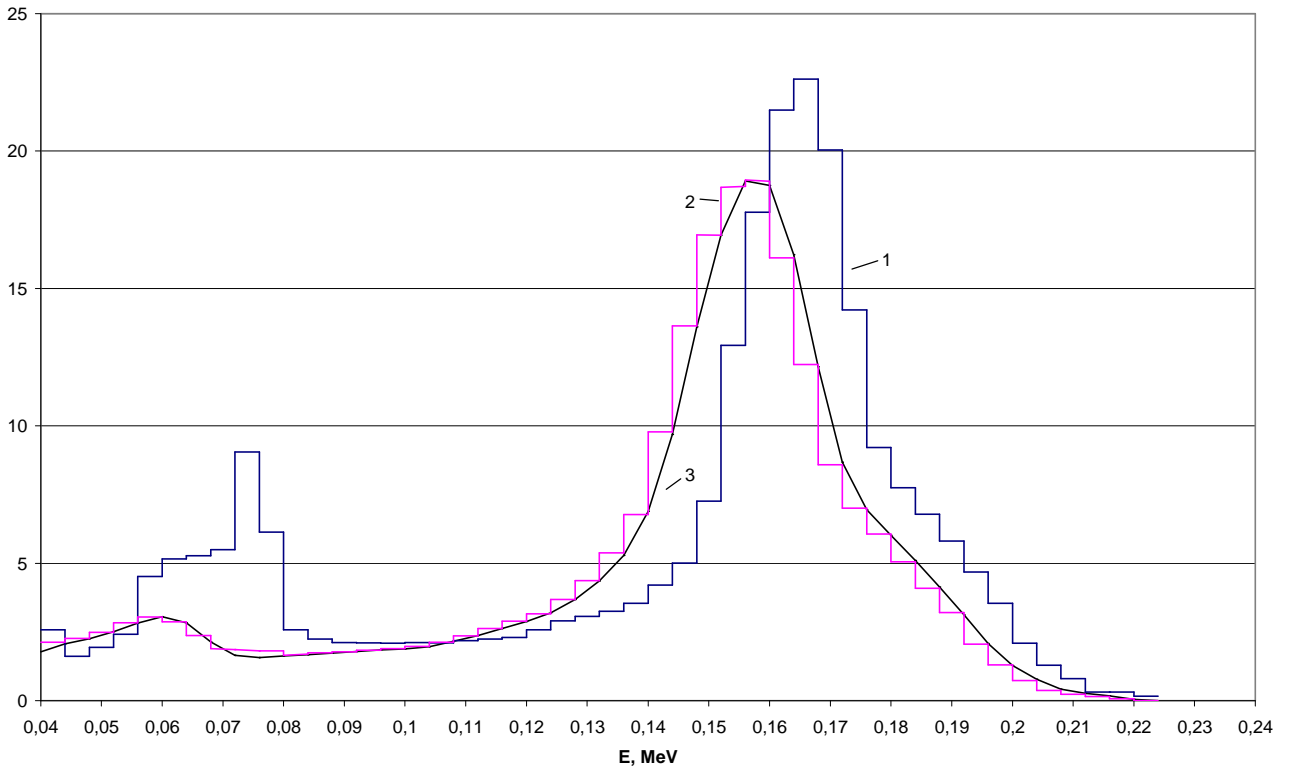

Figure 7. Electron energy spectra: 1 – after passing the grid, 2 – after passing the Al foil 13 μm thick; 3 – spectrum calculation according to ELIZA program based on the database [12] for each spectrum 1. The spectrum is normalized by the standard.

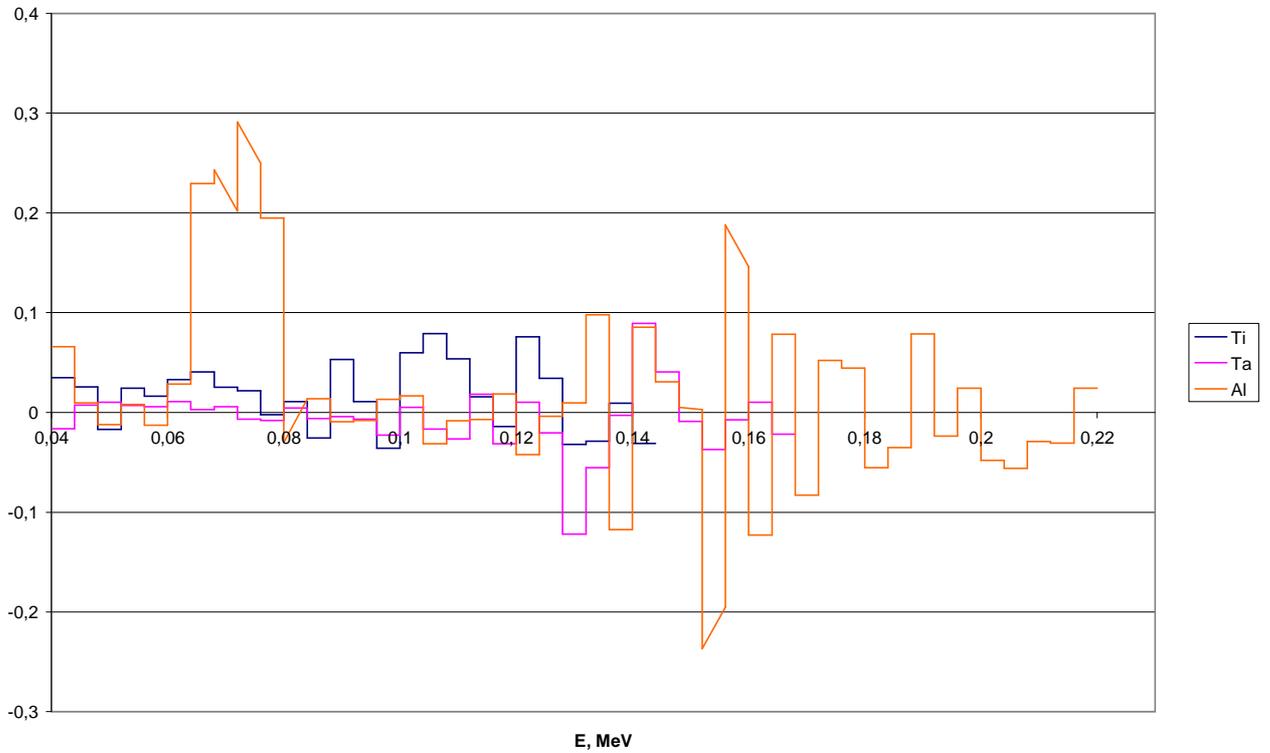

Figure 8. Difference of spectral density for theoretical and experimental spectra of electrons passed through Ti, Ta and Al foils.

### 3.2. Micropinch plasma electron gravitational radiation in pulse high-current discharges

The concept of a thermonuclear reactor on the principle of compressing dense high-temperature plasma by emitted gravitational field is supported by the processes of micropinching multicharged ion plasma in pulse high-current diodes. Figs 9, 10(a) show characteristic parts of micropinch soft X-ray radiation spectrum. Micropinch spectrum line widening does not correspond to existing electromagnetic conceptions but corresponds to such plasma thermodynamic states which can only be obtained with the help of compression by gravitational field, radiation flashes of which takes place during plasma thermalization in a discharge local space. Such statement is based on the comparison of experimental and expected parts of the spectrum shown in Fig. 10(a,b). Adjustment of the expected spectrum portion to the experimental one [15] was made by selecting average values of density ρ, electron temperature $T_e$ and velocity gradient U of the substance hydrodynamic motion.

As a mechanism of spectrum lines widening, a Doppler, radiation and impact widening were considered. Such adjustment according to said widening mechanisms does not lead to complete reproduction of the registered part of the micropinch radiation spectrum. This is the evidence (under the condition of independent conformation of the macroscopic parameters adjustment) of additional widening mechanism existence due to electron excited states and corresponding gravitational radiation spectrum part already not having clearly expressed lines because of energy transfer in the spectrum to the long-wave area.

That is to say that the additional mechanism of spectral lines widening of the characteristic electromagnetic radiation of multiple-charge ions (in the conditions of plasma compression by radiated gravitational field) is the only and unequivocal way of quenching electrons excited states at the radiating energy levels of ions and exciting these levels by gravitational radiation at resonance frequencies. *Such increase in probability of ion transitions in other states results in additional spectral lines widening of the characteristic radiation.* The reason for quick degradation of micropinches in various pulse high-currency discharges with multiple-charge ions is also clear. There is only partial thermolization of accelerated plasma with the power of gravitational radiation not sufficient for maintaining steady states.

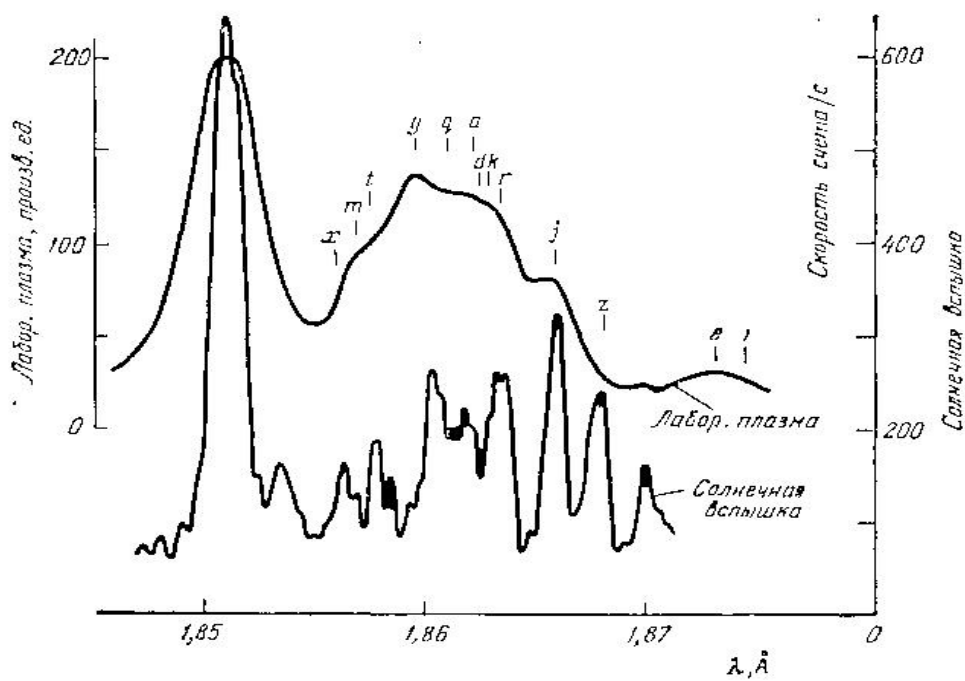

Figure 9. A part of vacuum sparkle spectrum and a corresponding part of solar flare spectrum. [14].

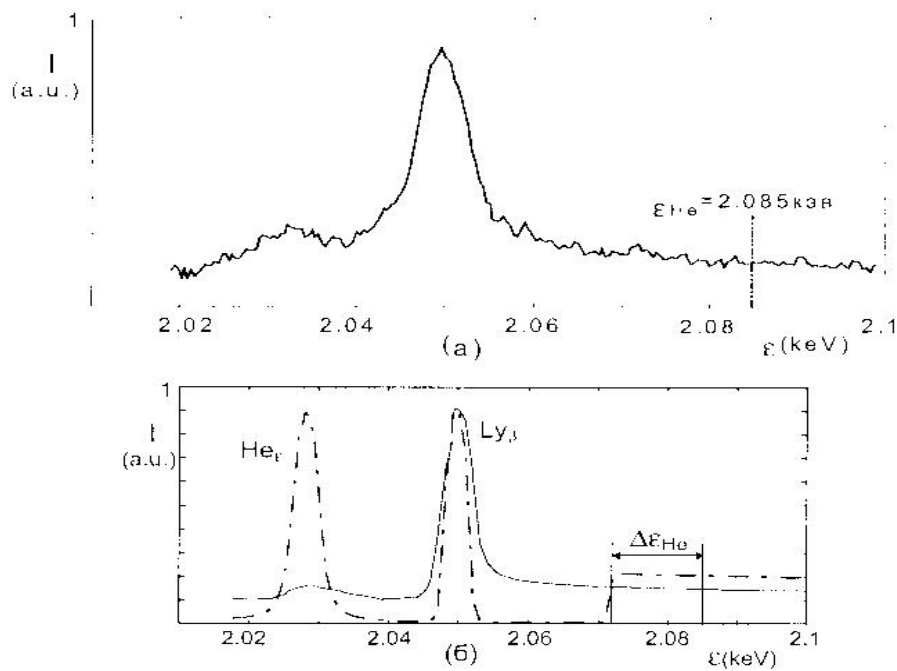

Figure 10. Experimental (a) and calculated (b) parts of a micropinch spectrum normalized for line Lyβ intensity in the area of the basic state ionization threshold of He-like ions.
The firm line in variant (b) corresponds to density of 0.1 g/cm$^3$ , the dotted line – to 0.01 g/cm$^3$; it was assumed that $T_e$ = 0.35 keV, [15].

It should be noted that widening of the spectral characteristic of the electromagnetic radiation [16] gave rise to a wrong conclusion about abnormal increase of the energy conservation law, while the widening of the spectral lines has a distinct and explicable nature, as stated above.

### 3.3. Thermonuclear plasma steady states generation

Available experimental data show that they can be reproduced in an active experiment directly with thermonuclear plasma on the ground of existence of gravitational radiation narrow-band spectrum in the range up to 171 keV with long-wave spectrum part realized by cascade electron transitions from the upper energy levels. Quenching lower excited states of electrons on the electron shell energy levels of heavy component ions in combination with cascade transitions will result in plasma compression in the conditions of blockage and gravitational radiation intensification.

Operation capacity of the chamber according to the scheme shown on Fig.3 with deuterium-tritium composition was tested experimentally. The obtained experimental data of plasma compression in the chamber [17]; however the holding time is not sufficient, there need to be longer holding time. The choice of such design as a design for a thermonuclear reactor is unequivocal since it is completely corresponds to the system of exciting and amplifying gravitational radiation when plasma is thermolized after outflow from the nozzle, and required additional compression actually takes place when the working plasma composition is changed (many-electron ions) and current-voltage characteristic of the charge changes correspondingly.

**References:**

1. Siravam C. and Sinha K., Phys. Rep. 51 (1979) 112.
2. Fisenko S. et all, Phys. Lett. A, 148,8,9 (1990) 405.
3. Ivanenko D.D. et al. Guage Theory of Gravity, Moscow, MGU Publishing House (1985).
4. Landau L.D. and Lifshitz E.M., Field Theory, Moscow, Publishing House «Nauka» (1976).
5. Warshalovich D.A. et al. Quantum Theory of Angular Momentum, Leningrad, Publishing House «Nauka» (1975).
6. Kamke E. Hand-book on Ordinary Differential Equations, Moscow, Publishing House «Nauka» (1976).
7. W. Pauli: Theory of Relativity; Pergamon Press; 1958.
8. Wu Z. S., Moshkovsky S. A., β-Decay, Atomizdat, Moscow (1970).
9. Fisenko S., Fisenko I., PCT Gazette № 46(2005) 553 (IPN WO2005/109970A1).
10. J. Robson, ibid, p.311;
11. C. Christensen et al. Phys. Rew. D5, 1628 (1972);
12. V. F. Tarasenko, S. I. Yakovenko, http://zhurnal.ape.relarn.ru/articles/2006/146.pdf
13. D. E. Cullen et al., Report IAEA – NDS-158, September (1994);
14. E.Ya. Goltz, I.A. Zhitnik, E.Ya. Kononov, S.L. Mandelshtam, Yu.V. Sidelnikov, DAN USSR, Ser. Phys., 1975, V. 220, p. 560);
15. V.Yu. Politov, A.V. Potapov, L.V. Antonova, Proceeding of International Conference "V Zababakhin Scientific Proceedings" (1998);
16. M.G. Haines et al., PRL, 96, 075003 (2006);
17. R. Lindemuth at al., PRL, v.75, #10, p1953 (1995).